\documentclass[a4paper,12pt]{article}
\usepackage[numbers]{natbib}
\usepackage{amsmath}
\usepackage{here}
\usepackage{amsthm}
\usepackage{graphicx,color}
\usepackage{ascmac}
\usepackage{amsfonts}
\usepackage{amssymb}
\usepackage{graphicx}
\usepackage{tikz}
\usetikzlibrary{intersections,calc,arrows.meta}
\usepackage{tree}
\usepackage{latexsym}

\usepackage{breqn}
\usepackage{mathtools}
\usepackage{authblk}
\begin{document}
	\title{Influence of trip distance and population density on intra-city mobility patterns in Tokyo during COVID-19 pandemic \thanks{We are grateful to all seminar and conference participants at the XV World Conference of Spatial Econometrics Association (SEA 2021), The 11th Asian Conference In Regional Science, and the 35th Applied Regional Science Conference for their useful comments and discussions. 
	Keisuke Kawata, Ștefana Cioban, Dao-Zhi Zeng, Tatsuhito Kono, and Tomokatsu Onaga also gave us valuable comments.
	This work was supported by the Research Institute for Mathematical Sciences, International Joint Usage/Research Center located at Kyoto University and the Open-type Professional Development Program (2021-SHIKOIN-7007). 
	N.F. was supported by JSPS KAKENHI Grant Number JP18K11462 and Promoting Grants for Research Toward Resilient Society, 2021.
	R.I. was supported by JSPS KAKENHI Grant Number JP18K01561.}}
	
	\author[1,\thanks{Corresponding author. E-mail: \tt{tsuboi@se.is.tohoku.ac.jp}}]{Kazufumi Tsuboi }
	\author[1,2,3,4,5]{Naoya Fujiwara}
	\author[1]{Ryo Itoh}
	
	\affil[1]{\it\scriptsize{Graduate School of Information Sciences, Tohoku University, Sendai, Miyagi, Japan.}}
	\affil[2]{\it Institute of Industrial Science, The University of Tokyo, Meguro, Tokyo, Japan.}
	\affil[3]{\it Center for Spatial Information Science, The University of Tokyo, Kashiwa, Chiba, Japan.}
	\affil[4]{\it Service Research Division, Tough Cyberphysical AI Research Center, Tohoku University, Sendai, Miyagi, Japan.}
	\affil[5]{\it PRESTO, Japan Science and Technology Agency, Kawaguchi, Saitama, Japan.}
	\date{December 22, 2021}
	\maketitle
	\begin{abstract}
	This study investigates the influence of infection cases of COVID-19 and two non-compulsory lockdowns on human mobility within the Tokyo metropolitan area. Using the data of hourly staying population in each 500m$\times$500m cell and their city-level residency, we show that long-distance trips or trips to crowded places decrease significantly when infection cases increase. The same result holds for the two lockdowns, although the second lockdown was less effective. Hence, Japanese non-compulsory lockdowns influence mobility in a similar way to the increase in infection cases. This means that they are accepted as alarm triggers for people who are at risk of contracting COVID-19.
    \end{abstract}
 \newpage
	\section{Introduction}
	Allegedly originating from Wuhan, Hubei province, China, the novel coronavirus disease (COVID-19) has prevailed all over the world to infect more than 400 million people and killed more than 4 million people as of August 2021 according to \cite{who2020}. The pandemic has changed our daily lives drastically.
	We maintain social distance and wear masks to save ourselves from infection.
	People stay home, work online; all these changes significantly reduce our daily mobility within cities. 
	
	Governments in many countries have also conducted non-pharmaceutical interventions (NPIs), including lockdown policies.
	NPIs have ordered or asked people to stay at home, encouraged onsite workers to start teleworking, and closed schools and public facilities for several weeks according to \cite{hale2021global}. 
	The type of NPIs vary across countries; most countries and cities imposed compulsory lockdowns with penalties for non-compliance, while some countries and cities, such as Sweden, Japan, and New York, imposed almost non-compulsory ones as summarized in \cite{born2021lockdown}. 
	Compulsory lockdowns are usually known to be more effective in mitigating infections than non-compulsory lockdowns \cite{megarbane2021lockdown}. 
	However, there are concerns that long and strong restrictions cause a significant loss of income and opportunities, which can decrease people’s welfare \cite{acemoglu2020multi,bonaccorsi2020economic}. 
	Therefore, we need to investigate how infection risks and lockdowns affect mobility in daily life. 
	
	Many previous studies have examined the negative impact of COVID-19 and lockdowns on human mobility, which is promoted by the recent growing accessibility to various mobile phone data ~\cite{glaeser2020jue,kraemer2020effect}. 
	There is a positive relationship between mobility inflow to each county and the number of infections in the United States using mobile phone data\cite{xiong2020mobile}.
	A previous study  quantified social contacts and human mobility using publicly available smartphone GPS data and Place IQ \cite{couture2021jue}.
	It was also shown by \cite{schlosser2020covid} that Lockdowns not only reduce mobility but also cause structural changes in mobility, such that many people quit long-distance travel and their networks become smaller.
	
	The differences between compulsory and non-compulsory lockdowns have also been examined. 
	Ref~\cite{megarbane2021lockdown} compared various countries’ lockdowns and revealed that countries conducting compulsory lockdowns reduce infection cases and recover more rapidly than non-compulsory lockdowns such as in the United States and Sweden. 
	They claimed that mobility restrictions were particularly effective in the early stages of an outbreak.
	Ref~\cite{tobias2020evaluation} revealed the relationship between lockdowns and infection cases in Italy and Spain.\footnote{In the northern parts of Italy, mobility was restricted from March 8, 2020. On March 10, this restriction was imposed throughout the country. Italy implemented two stages of lockdowns. The slope of the increasing rate in daily diagnosed cases, daily deaths, and ICU daily admissions were more stable during the first lockdown, but infection trends continued to rise. After that, a more restrictive lockdown was implemented on March 21. All businesses were closed, except for the essential industries, and the increasing trends of COVID-19 infections changed. Spain imposed similar lockdowns and yielded similar results. }
	In contrast, several studies show that the first Japanese non-compulsory lockdowns implemented in the spring of 2020 were effective \cite{watanabe2021japan,yabe2020non,hosono2021epidemic}.
	\footnote{
		Ref~\cite{watanabe2021japan} identified the effects of the first Japanese lockdown into the intervention effect and the information effect considering that only the information effect appears in the neighboring prefectures of the prefectures which implemented the voluntary lockdowns. 
		Ref~\cite{hosono2021epidemic} investigated the effects of infection risks and lockdowns on people and their choice to stay at home. 
		Ref~\cite{yabe2020non} quantified the mobility reductions after the first lockdown and showed that human mobility decreased by approximately 50\%, and this reduction contributed to a 70\% decrease in social contact. They used cellphone data; specifically, 
		ref~\cite{watanabe2021japan} and  
		\cite{hosono2021epidemic} used identical data sources. }
	Ref~\cite{watanabe2021japan} also showed that the first Japanese lockdown had a huge information effect in allowing people to know the infection risk and refrain from going out. 
	
	However, most previous studies did not use spatial characteristics of daily trips, such as distance and population density of destinations, in considering the effect of voluntary lockdowns. 
	Although some studies used cellphone data, they only investigated macro-level data such as intercity trips or aggregated trips at the city level. 
	Hence, there are only a few implications about detailed spatial variations in the effects of voluntary lockdowns and infection risks, although they are useful for understanding how and why voluntary lockdowns matter in terms of  daily trips. 
	Furthermore, changes in people’s attitudes over one year during the pandemic have not been revealed so far. 
	
	This study aimed to reveal how human mobility, especially commuting, was affected by the infection risks of COVID-19 and the lockdown policy in the Tokyo metropolitan area (hereafter, Tokyo MA) in Japan. 
	The novelty of this study is twofold. 
	First, we focus on several characteristics of trips related to infection risks, such as population density and distance, and the variation in the effects of voluntary lockdowns as well as infection risks on daily, intra-city trips by these factors. 
	Our data, collected and estimated from cellphones, are suitable for this purpose as they capture people’s hourly location at 500m $\times$ 500m cell level with their residency. 
	Using the data, we show the effects of infection risks, captured by the number of new infection cases, and that the implementation of lockdowns is more significant if the population density of the destination is large and the distance of the trip is long. 
	This result implies that implementation of lockdowns made people more sensitive to the infection risks; hence, we could show new evidence for the information effect of the policy. 
	Second, our data period was from January 2020 to March 2021. This is a little longer than previous studies and is long enough to capture the change in people’s attitudes toward infection risks and the lockdowns over more than a year.
	We show that the effects of both infection cases and lockdowns became smaller in the latter half of our data period.

	\section{Background}
	\subsection{Summary of COVID-19 in Tokyo}
	The outbreak of COVID-19 was first noticed in Wuhan, Hubei Province, China, and rapidly prevailed worldwide. 
	In Japan, the first infection was confirmed on January 15, 2020. After the first case, the infection spread gradually.
	
\begin{figure}[H]
	\centering
	\includegraphics[width=1.0\linewidth]{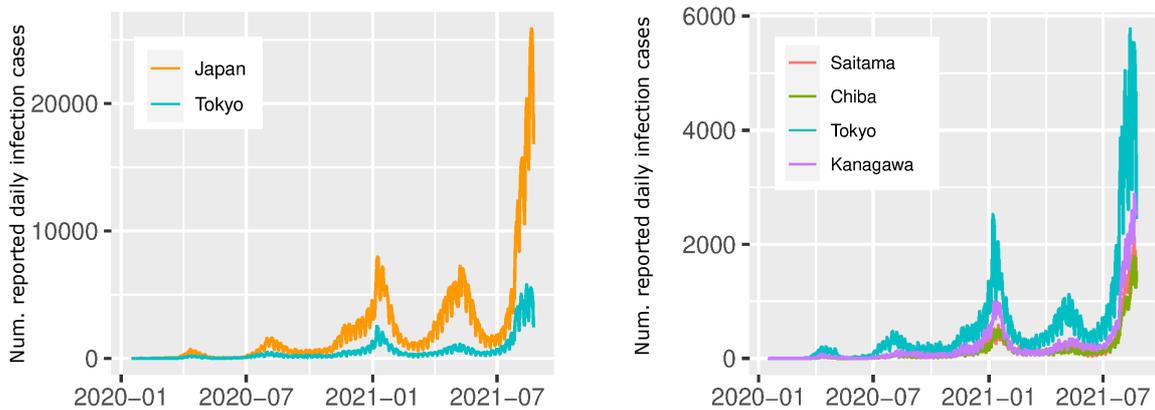}
	\caption{{\bf Infection cases in entire Japan (A) and  Tokoy MA (B).}}
	\label{fig1}
\end{figure}

	Figure 1 shows the cases of infection in Tokyo and Japan. Both graphs have several waves and exhibit similar trends. In Fig~\ref{fig1}\,(A), we focus on the Tokyo MA. Fig~\ref{fig1}\,(B) shows the number of cases in different prefectures in Tokyo MA. The city has a similar trend to infection cases. In Tokyo, 2,520 people were infected with COVID-19 on January 7, 2021. 
	Further, infection cases increased again around the Olympic Games in Tokyo. 
	
	\subsection{Review of lockdowns in Japan and other countries}
	Many countries have implemented lockdowns in various ways. 
	In this subsection, we show the features of Japanese lockdowns compared to those of other countries.
	Most countries such as China and Spain imposed compulsory lockdowns and implemented penalties for non-conformists.
	\footnote{China imposed strict compulsory lockdowns. In Wuhan, all public transportation via airports, stations, and some roads, was suspended on January 23, 2020. Moreover, the government restricted the population movement. On March 10, 2020, the Chinese government controlled individual activity more strictly  using smartphone applications and QR codes, and Chinese people had to register their mobility.}
	However, some countries such as Japan, Sweden, and the United States implemented voluntary lockdowns merely by asking people to refrain from going out.
	
	The United States implemented different NPIs across states, and non-compulsory policies were implemented in some cities such as New York, in which the government announced a state of emergency on March 9, 2020.
	The government did not restrict private activities such as shopping and going out, but ordered residents not to commute to their office, except for essential workers in health care, social assistance, and the public administration industry.
	Any company violating this law was fined.
	In contrast, Japanese and Swedish NPIs asked people to refrain from going out, but were not directly enforceable on individuals. Residents’ mobility was not restricted, and the government did not impose penalties. We call such a policy ``quasi lockdown" in this paper.
	
	Japan experienced two quasi-lockdowns during our data period, from January 2020 to March 2021.
	\footnote{As of July 31, we know that two more lockdowns have been implemented. The third was from April 25, 2021, but the affected areas were limited. The end of the periods differ across prefectures. The fourth lockdown started from August 1, 2021, as infection cases increased during the Tokyo Olympics.}
	The first quasi-lockdown was from April 7 to May 25, 2020. From April 7, eight prefectures, including Tokyo, Kanagawa, Chiba, and Saitama prefectures, imposed a quasi-lockdown. From April 16, this lockdown spread nationwide. This national lockdown ended on May 14, except in Tokyo, Kanagawa, Chiba, Saitama, and Hokkaido prefectures. The lockdown in Tokyo, Kanagawa, Chiba, and Saitama ended on May 25. The second quasi-lockdown continued from January 8, 2021, in Tokyo, Kanagawa, Chiba, and Saitama prefectures to March 21, 2021. During the second quasi-lockdown, other prefectures were added and eliminated.
	
	There are four key features of Japanese lockdowns. First, the Japanese government set the goal of reducing onsite workers by $ 70\% $, and many workers started to telecommute. Second, many restaurants with entertainment such as night clubs, hosts, and hostess bars were restricted from opening. If restaurants cooperated with the regulations, they received compensation payments from the government.
	\footnote{ In the second quasi-lockdown, the government gave more detailed designations regarding restaurants’ opening time. Restaurants were requested to close by 8 PM and serving alcoholic drinks was limited from 11 AM to 7 PM.}
	In contrast, some libraries, museums, and parks were excluded from this policy. Third, the headcount for large events was at most 5,000 people or $ 50\% $ of the total capacity. Fourth, the Japanese government set 0.5 infection cases per 100,000 people as a decision criterion for canceling the quasi-lockdown. 
	
	These facts show that no restrictions were imposed directly on the individuals. Although a few restrictions were imposed on some public facilities and the opening of restaurants at night, their effects on daytime mobility, such as commuting, are considered quite indirect and limited. 
	
	In this study, we used four prefectures’ data in Tokyo MA: Tokyo, Kanagawa, Chiba, and Saitama prefectures, because the infection trend is similar for each prefecture and the period of quasi-lockdown was the same.
	
	\section{Data}
	\subsection{Data}
	Three types of data were used in the present analysis. The first is mobility data and mobile spatial statistics data collected from cellular phones.
	The second is infection data which record the prefecture-level infection cases.
	The third is geographical data, used to calculate the origin-destination (OD) matrix.
	
	\subsubsection{Mobile spatial statistics data}
	We use mobile spatial statistics data provided by DOCOMO Insight Marketing, INC.
	The data are the estimated hourly population in grid cells of approximately 0.25 $ km^2 $ based on the location information.
	The database was collected from the cellular phone users of NTT DOCOMO, INC., one of the largest cellular phone operators in Japan.
	Mobility data are an expanded estimate of the number of populations from cellular phone users. This database eliminates some grid cells with fewer than ten people to protect users’ privacy.
	
	Mobility data also include information about users’ residential cities; hence, the number of people who live in city $i$ and stay at cell $j$ at every hour is available. This study uses the population as the number of trips from city $i$ to the grid cell $j$.
	However, if the grid cell $j$ is included in city $i$, the remaining population may include people who stay home, so this study excludes such samples from our analysis.
	
	\subsubsection*{The number of COVID-19 cases}
	We use data from the daily reported new cases of COVID-19 aggregated by NHK\cite{nhk2021}, the public broadcaster in Japan. This dataset counts new infection cases by prefecture, which are frequently broadcasted as indicators of spreading COVID-19.
	We use them for independent variables to explain mobility data.
	\subsubsection*{Geographical data}
	We calculated the distance of trips between the center of the grid cells and the residence of samples.
	We used the location of the city hall of the residential cities instead of the exact residential location. 
	We also employed the Euclidean distance as a proxy of trip distance.
	\footnote{Location information of city halls is obtained from the Digital National Land Information (https://nlftp.mlit.go.jp/ksj/jpgis/datalist/KsjTmplt-P05.html). }
	All the location data are included in the QGIS~\cite{qgis2021}, and the distances are calculated using R.
	\footnote{We used the distGeo package in R that calculates geodesic distance on an ellipsoid.}
	
	\subsubsection*{The data period and targeted area}
	The data period used in our analysis was from January 2020 to March 2021, which includes two lockdowns. 
	Although the original dataset covers over entire Japan, this study uses limited samples in Tokyo MA.
	
	First, our analysis mainly used the population location at 10:00 a.m.
	Further, to control for the effect of days, we used the data from every Thursday, except for national holidays. 
	Although the objectives of trips are not available in our data, we are especially interested in commuting trips and it is a typical time zone to capture such trips.  
	However, we also examined the mobility on holiday afternoons, which is considered mixture of variety of trips. 
	
	Second, we only used Tokyo MA because this area experienced two significant lockdowns in our data period.
	\footnote{In this study, Tokyo MA is defined to comprise four prefectures: Tokyo, Kanagawa, Chiba, and Saitama. It is different from the definition of the Ministry of Land, Infrastructure, Transport and Tourism (MLIT) based on commuting, which includes parts of other prefectures, while it is not part of those four prefectures.}
	Moreover, since Tokyo MA is the largest metropolitan area in Japan with more than 36 million people, there are a sufficient number of destination grid cells with high population density. 
	Therefore, our dataset includes 60 time points, 28,815 grid cells for destinations of trip, and 244 cities for origins.
	
	\subsection{Summary of the mobility data}
	\subsubsection*{Patterns of morning trips}
\begin{figure}[H]
		\centering
		\includegraphics[width=1.0\linewidth]{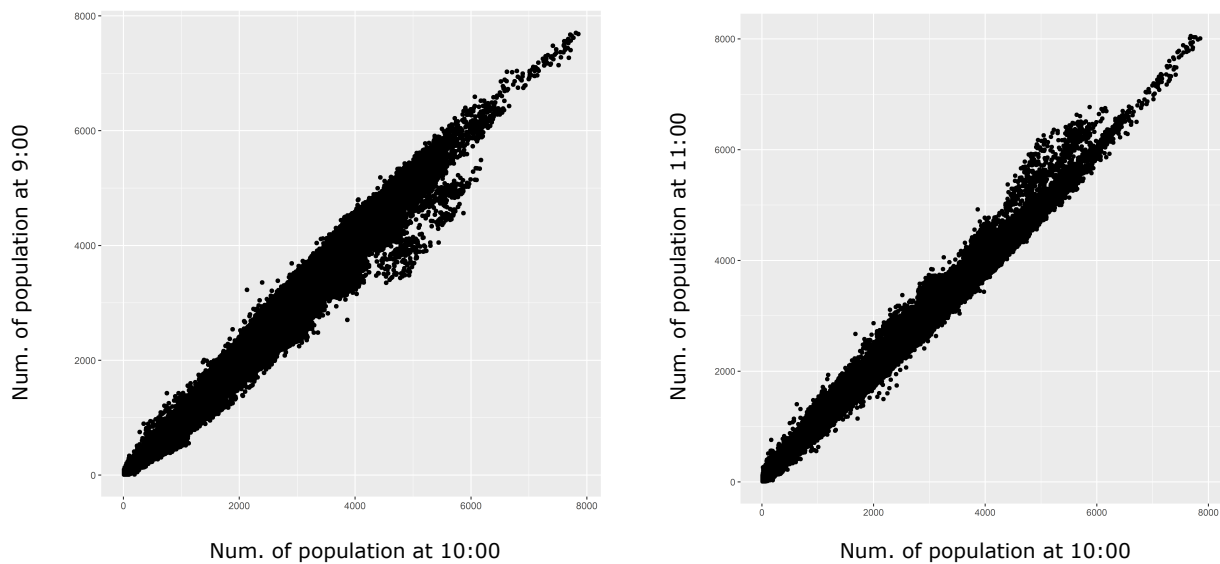}
		\caption{{\bf comparison of staying population at 10:00 a.m. to 9:00 and 11:00 a.m.} The data includes the average population on Thursdays in November 2019 in Tokyo MA.}
		\label{fig2}
	\end{figure}

	Our analysis mostly uses the location data of 10:00 a.m. based on the notion that it is the best to capture commuting. 
	However, to check the reasonability of the idea, we also show the locations at 9:00 a.m. and 11:00 a.m.
	Fig~\ref{fig2} shows the strong correlation of the number of trips at 9:00 a.m. and 11:00 a.m. with that at 10:00 a.m., which means that the location is almost stable from 9:00 to 11:00 a.m.
	We can anticipate from the stability that commuting almost finishes by 10:00 a.m., and trips for other objectives are still few in the morning. 
	Therefore, it is supposed that the population location at 10:00 a.m. is mostly the workplace, including students’ schools, hence the data at 10:00 a.m. is suitable to capture commuting trips. 
	
	\subsubsection*{The effect of quasi lockdowns}
	We intend to examine the effect of lockdowns. Hence, we will provide an overview of the data to show how human mobility changes during lockdowns.
	\begin{figure}[H]
		\centering
		\includegraphics[width=0.7\linewidth]{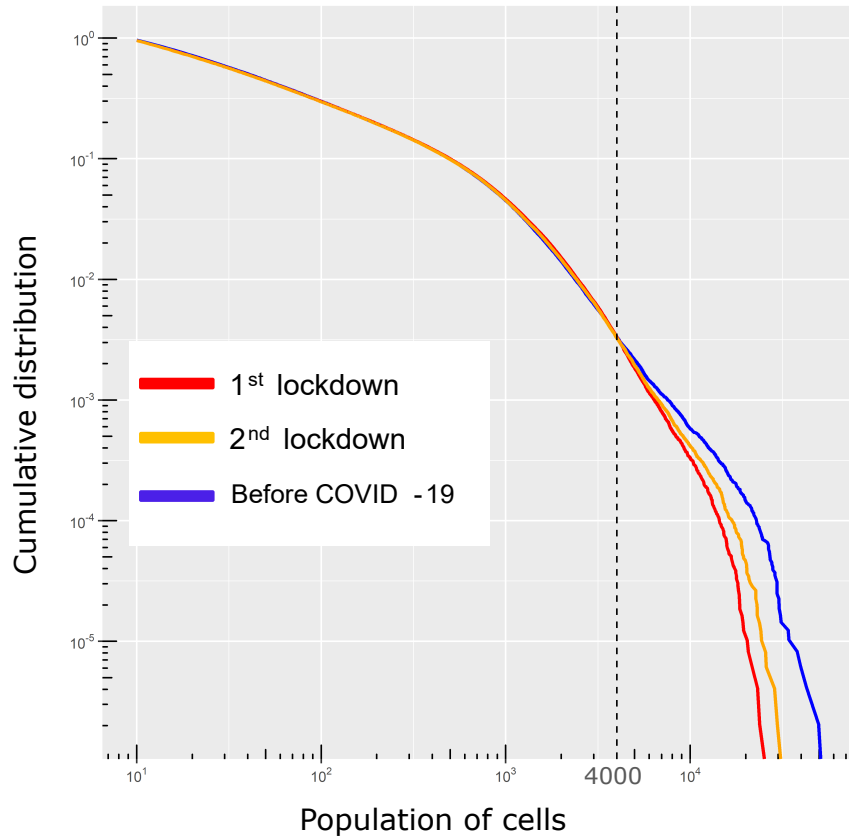}
		\caption{{\bf Complementary cumulative distribution function of population in approximately 500m $ \times $ 500m grid cells.}}
		\label{fig3}
	\end{figure}
	 \begin{figure}[H]
	 	\centering
	 	\includegraphics[width=0.7\linewidth]{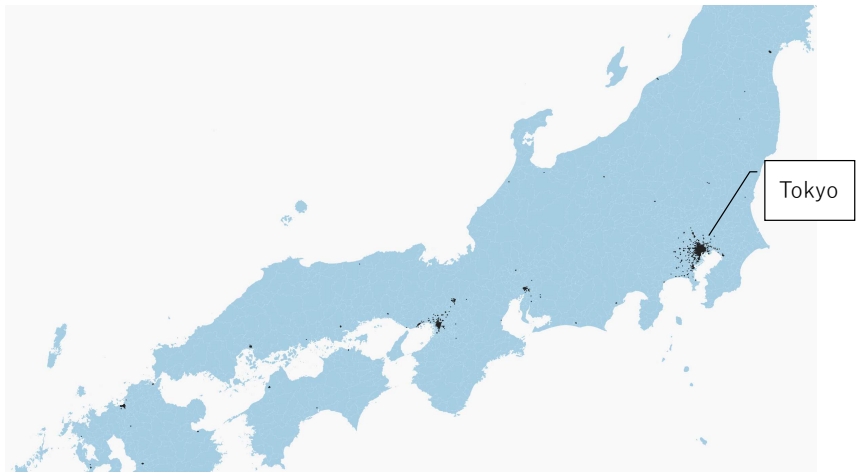}
	 	\caption{{\bf Location of high-density places.}Black dots in the map represent grid cells in which the average population at 10:00 a.m. on Thursdays in November 2019 exceeded 4,000.}
	 	\label{fig4}
	 \end{figure}

	First, we show how the significance of lockdown effects differs by population density. Fig~\ref{fig3} shows the complementary cumulative distribution of the population in each cell.  
	The blue line indicates the population in each cell at 10:00 a.m. before COVID-19 (November 7, 2019). The red and yellow lines indicate the population during the first (April 9, 2020) and second (March 11, 2021) quasi-lockdowns, respectively. All three days are Thursdays, except for holidays. Dates of both red and yellow lines were chosen for the first Thursday during the lockdown.
	Fig~\ref{fig3} shows that significant change in mobility occurs in grid cells with a population larger than 4,000. Although the share of grid cells with more than 4,000 people in all the populated grid cells is less than 1\%, they share 10.4\% of the total population in Japan; hence the lockdowns influenced a large number of people.
	
	Further, the black points in Fig~\ref{fig4} show the locations of highly populated places. Most of these points are concentrated in large metropolitan areas such as Osaka and Chubu, and especially Tokyo. This research, therefore, focuses on the Tokyo MA because it experienced two lockdowns and had a sufficient number of sample grid cells affected by them. 
	
\begin{figure}[H]
		\begin{center}
			\begin{tabular}{c}
				
				\begin{minipage}{0.5\hsize}
					\begin{center}
						\includegraphics[clip, width=80mm]{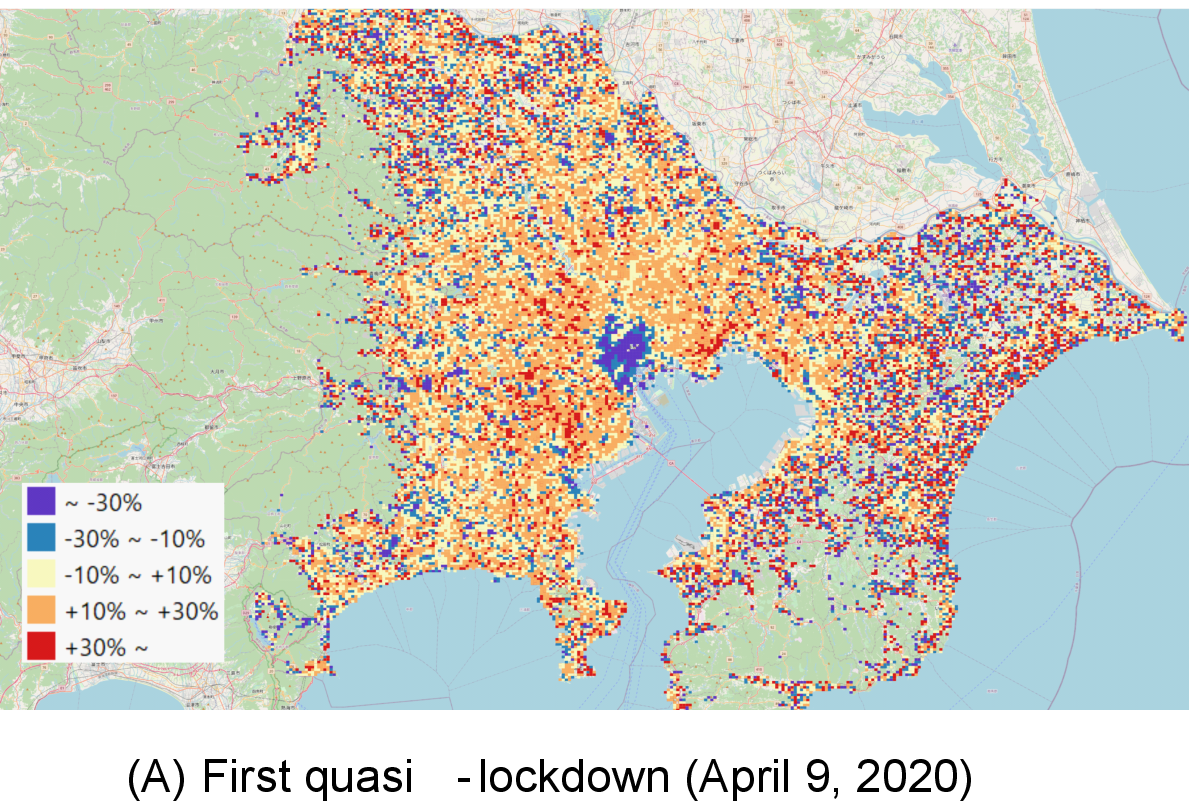}
					\end{center}
				\end{minipage}
				
				\begin{minipage}{0.5\hsize}
					\begin{center}
						\includegraphics[clip, width=80mm]{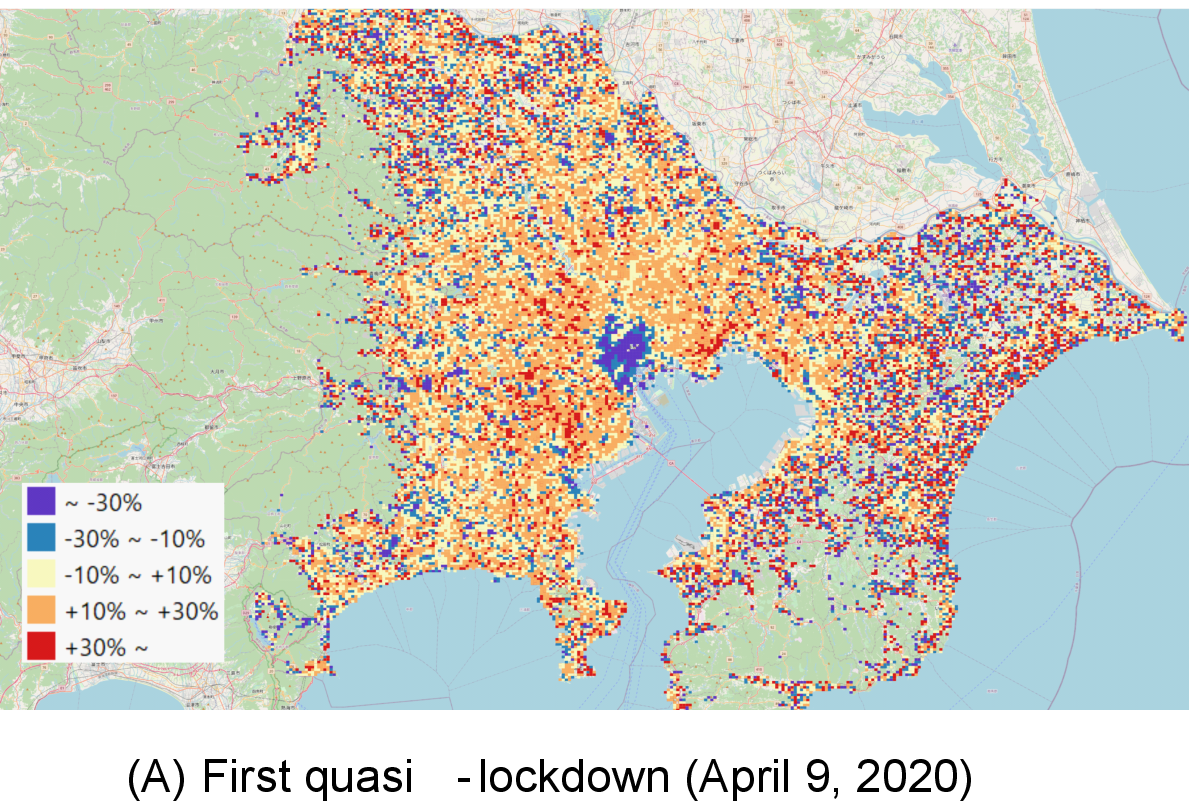}
					\end{center}
				\end{minipage}
				
			\end{tabular}
			\caption{{\bf Population density in Tokyo MA in comparison to before COVID-19: (A)First lockdown; (B) Second lockdown.}}
			\label{fig5}
		\end{center}
	\end{figure}

	Fig~\ref{fig5} visualizes the change in staying population by the lockdowns in the Tokyo MA by comparing the average population of each cell between lockdown periods and before COVID-19.
	\footnote{We took the average number of people every Thursday from November 7 to December 19, 2019. We omitted December 26 from the average population.}
	In the central area of Tokyo, mobility decreases by more than 30 \%, while the staying population in the surrounding areas increases as a result of the stay home movement. They also show that the effect of the first lockdown is more significant than that of the second one.

	\section{Estimation and Results}
	We used panel data to conduct a fixed-effect estimation. The equation to estimate is
	
	\begin{eqnarray}
	\label{eq:estimation}
	\ln P_{ij,t}=\alpha_{ij}+\alpha_t +\beta_1 \mathbf{X}_{ij,t},
	\end{eqnarray}
	where $ P_{ij,t} $ is the number of people moving from residential city $j$ to the destination grid cell, $i$, on day $t$. Since our primary interest is commuting behavior, the data at 10:00 a.m. on every Thursday, except for national holidays, is used. $\alpha_{ij}$ is the fixed effect determined by the OD pair, $i$, and $j$. $\alpha_t$ is the fixed effect of each date $t$. 
	
	$\mathbf{X}_{ij,t}$ is the vector of the independent variables as follows: First, we used the number of infection cases from one week ago (IFC) in the destination prefecture to show infection risk by a trip. 
	We consider time lags of reaction for COVID-19 spreading in these indices. 
	For example, some onsite workers change their commuting patterns in the next week after watching news about COVID-19 on weekends. 
	\footnote{To capture this time lag, we use data from Monday to Sunday in weekday analysis, and from Sunday to Saturday in weekend analysis.}
	Following the same idea, we also used the number of infection cases two weeks prior to that of one week ago (LONG) to describe how the recent (short-term) outbreak is prolonged.
	If this index is large, people will have enough time to change their behavior accordingly.
	
	Second, $\mathbf{X}_{ij,t}$ also includes the lockdown dummy variables that take 1 during each of the first and second lockdowns in the Tokyo MA; they are implemented from April 7 to May 25, 2020, and from January 8 to March 25, 2021, respectively. 
	
	Third, it also includes the cross terms of the above variables, the number of infections and lockdown dummies, the population density of destination cells, and trip distance.
	\footnote{Detailed calculation is given in Section 3.1.}
	These cross terms capture different reactions to the virus, prevailing by the characteristics of trips. We use the population in the sample grid cell $i$ before COVID-19 as the population density. Single terms such as distance, density, and lockdown dummies in few results are excluded because they are included in the fixed effects of OD and date.
	
	\subsection{Result of the baseline estimation}
	Our panel estimations of Eq~(\ref{eq:estimation}) are reported in Table~\ref{table1}. 
	\begin{table}[H]
			\centering
			\caption{
				{\bf Baseline results}}
			\begin{tabular}{l c c c c}
				\hline
				& Model 1 & Model 2 & Model 3 & Model 4 \\
				\hline
				ln(IFC)            & $-0.0111^{***}$ & $0.0995^{***}$  & $-0.0474^{***}$ & $0.0749^{***}$  \\
				& $(0.0001)$      & $(0.0013)$      & $(0.0004)$      & $(0.0014)$      \\
				ln(LONG)           & $-0.0175^{***}$ & $0.0845^{***}$  & $-0.0342^{***}$ & $0.0988^{***}$  \\
				& $(0.0001)$      & $(0.0015)$      & $(0.0003)$      & $(0.0015)$      \\
				LD1                 & $-0.2197^{***}$ & $0.5892^{***}$  &                 &                 \\
				& $(0.0007)$      & $(0.0085)$      &                 &                 \\
				LD2                 & $-0.0005$       & $0.0741^{***}$  &                 &                 \\
				& $(0.0003)$      & $(0.0040)$      &                 &                 \\
				ln(IFC)$\times$ln(Dens)  &                 & $-0.0091^{***}$ &                 & $-0.0081^{***}$ \\
				&                 & $(0.0001)$      &                 & $(0.0001)$      \\
				ln(IFC)$\times$ln(Dist)  &                 & $-0.0041^{***}$ &                 & $-0.0043^{***}$ \\
				&                 & $(0.0002)$      &                 & $(0.0002)$      \\
				ln(LONG)$\times$ln(Dens) &                 & $-0.0108^{***}$ &                 & $-0.0108^{***}$ \\
				&                 & $(0.0001)$      &                 & $(0.0001)$      \\
				ln(LONG)$\times$ln(Dist) &                 & $-0.0016^{***}$ &                 & $-0.0036^{***}$ \\
				&                 & $(0.0002)$      &                 & $(0.0002)$      \\
				ln(Dens)$\times$LD1       &                 & $-0.0850^{***}$ &                 & $-0.0844^{***}$ \\
				&                 & $(0.0005)$      &                 & $(0.0005)$      \\
				ln(Dist)$\times$LD1       &                 & $-0.0159^{***}$ &                 & $-0.0158^{***}$ \\
				&                 & $(0.0011)$      &                 & $(0.0011)$      \\
				ln(Dens)$\times$LD2       &                 & $0.0000$        &                 & $-0.0060^{***}$ \\
				&                 & $(0.0003)$      &                 & $(0.0003)$      \\
				ln(Dist)$\times$LD2       &                 & $-0.0087^{***}$ &                 & $-0.0037^{***}$ \\
				&                 & $(0.0005)$      &                 & $(0.0005)$      \\
				\hline 
				FE : OD             & YES           & YES           & YES           & YES           \\
				FE : date           & NO            & NO            & YES           & YES           \\
				\hline
				$R^2$               & $0.93$        & $0.94$        & $0.94$        & $0.94$        \\
				Adj. $R^2$          & $0.93$        & $0.94$        & $0.94$        & $0.94$        \\
				within.r.squared    & $0.09$        & $0.13$        & $0.01$        & $0.05$        \\
				Observations                & $10,060,792$    & $10,060,792$    & $10,060,792$    & $10,060,792$    \\
				\hline
			\end{tabular}
			\begin{flushleft}
				Notes: Figures in parentheses are cluster-robust standard errors by OD. $^{***}$, $^{**}$, and $^{*}$ denote statistical significance at the 0.1\% , 1\%, and 5\% level, respectively.
			\end{flushleft}
			\label{table1}
	\end{table}
	Model 1 shows the result of the simple panel estimation without cross terms. 
	Both IFC and LONG were negatively correlated with mobility, showing that commuting decreased as infection cases increased. From the result, a 1\% increase in IFC decreases mobility by 0.011\%. Lockdown dummies LD1 and LD2 also have significant negative effects, but LD2 is less significant. Note that this analysis only considers short-run changes, but there will be some long-run effects if such increases in infection cases last for a long time because people may change their lifestyles fundamentally. Additionally, the implementation of the first and second lockdown decreased mobility by 22.2\% and 0.26\%, respectively.
	\footnote{Note that LDs are dummies taking the values 1 or 0 unlike IFC for which the natural logarithm is used for independent variable. Therefore, when LD1 changes from 0 to 1, the mobility changes by $\beta_{LD1} \times $100\%, where $ \beta_{LD1} $ is the coefficient of LD1.}
	This study also shows a large influence of the Japanese voluntary lockdown, as in former studies, as well as a statistically significant but weak effect of the second lockdown. 
	
	Model 2 includes cross terms for the two types of infection data and lockdown dummy variables for commuting distance and population density.
	All cross terms related to infection cases are negatively correlated with mobility. \footnote{Understanding these results may be complicated. We explain the cross terms using $ \ln(IFC) \times \ln(Dens) $ as an example. When we focus on one day, which means that we set $ \ln(IFC)$ fixed, then the negative coefficient shows that the higher population density grid cells have, the fewer people commute to such grid cells as the destination.} 
	This implies that people avoid long commutes and crowded places when infection cases increase.
	This behavior is rational because when the number of infected people increases, the risk of infection in crowded places where there is a lot of contact between people increases significantly.
	\footnote{The risk of infection is generally determined by the number of contacts with the infected people, which is positively correlated with the number of contacts with others and the share of the infected people in the population. This is why risk increases significantly in congested places when the number of infection cases increases.}
	This is also the case for long-distance trips where people are exposed to the risk in congested trains for a long time.
	\footnote{However, this result is also explained by the large incentive to introduce telecommuting for people who commute long distance. When the pandemic made teleworking more socially acceptable, such people were more likely to choose telework to save their large commuting costs.}  
	
	The cross terms related to the lockdowns also show negative coefficients for the first lockdown, while the effect of density is ambiguous for the second lockdown. That is, people avoid long-distance trips and crowded places when lockdowns are imposed; a similar reaction is seen when there is an increase in infection risks, described by the number of infection cases. This similarity can be explained by the information effect of lockdowns presented by \cite{watanabe2021japan}. The governments of Japan and Tokyo decided to impose this lockdown considering the increasing infections, and it was widely known through the media; hence, people accepted the lockdowns as warnings against the risk of infection. Although we do not identify the information effect from other effects such as the intervention, unlike \cite{watanabe2021japan}, our results provide additional evidence for the existence of the information effect of the voluntary lockdown. 
	
	Furthermore, one may consider that the coefficient of the single term of IFC is positive, and mobility might increase when the number of infection cases increases. However, infection also decreases the mobility via the cross terms, and the total marginal effect is negative, as examined in the later part. 
	
	Finally, we introduce time-fixed effects in Models 3 and 4, where the lockdown dummies must be omitted because they are included in the time-fixed effects. Despite the control of an additional fixed effect, most estimates do not change drastically; hence, most of our results are robust.

	\subsection{Total effects and their sensitivity}
	Our key variables, number of infections and lockdowns, affect mobility via multiple cross terms in Model 2. Since the total marginal effects differ depending on the distance of trips and density of destination, we need to determine how much the effect varies by them. In addition, every single coefficient does not simply tell us the total influence quantitatively when they change, and we need to evaluate these variables by their total effects. Therefore, we calculated the total effects and their sensitivities to density and distance. 
	
\begin{figure}[H]
		\centering
		\includegraphics[width=0.7\linewidth]{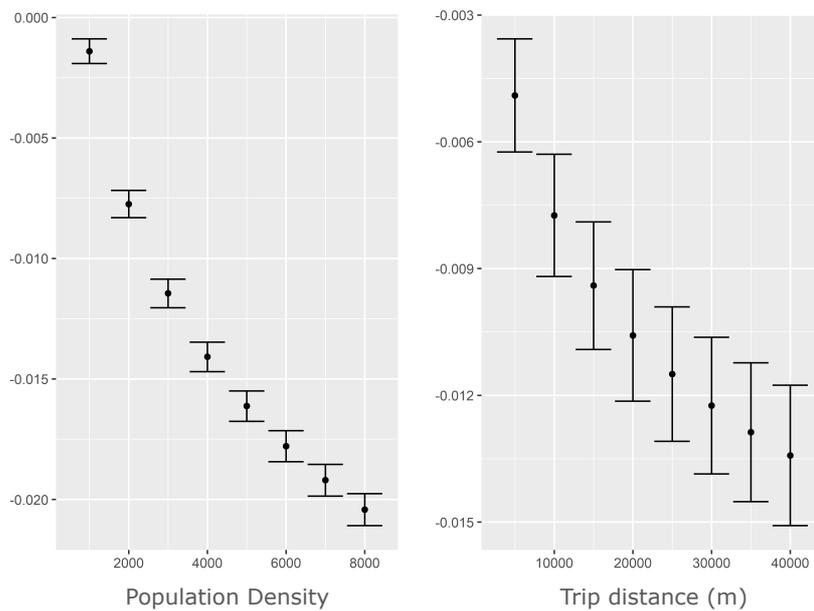}
		\caption{{\bf Total marginal effects of infection cases (IFC).} Population density and trip distance are fixed to 2000 and 10km when the other variable is changed.}
		\label{fig6}
	\end{figure}

	To quantify the total marginal effect of IFC, we define the elasticity of the mobility change as follows:
	\begin{eqnarray}
	e=\frac{\frac{d\ P}{P}}{\frac{d\ IFC}{IFC}}=\frac{d \ln{P}}{d \ln{IFC}}.
	\end{eqnarray}
	This elasticity describes the total effect of how many percent decrease of mobility does one percent increase of infection cases affect. From Eq (\ref{eq:estimation}), we can calculate $ e $ as follows:
	\begin{eqnarray}
	e=\beta_0+\beta_1\ln{Dist}+\beta_2\ln{Dens},
	\end{eqnarray}
	where $ \beta_0, \beta_1 $, and $ \beta_2 $ denote the coefficients of $ \ln IFC, \ln IFC \times \ln{Dens} $, and $ \ln{IFC} \times \ln{Dist} $ obtained from Model 2, respectively. This elasticity depends on the values of the distance and population density. Therefore, we use specific values to check whether mobility increases or decreases in specific areas.
	Fig~\ref{fig6} shows the total marginal effects of IFC, where dots represent the estimated values and bars represent 95\% confidence intervals. 
	They show that the total effect of IFC are almost negative. 
	Moreover, the total effects are highly sensitive to the population desnsity and trip distance. 
	Suppose the distance of the destination is fixed at 10 km and the population density is changed from 2,000 to 6,000; the total effect of infection cases ranges from -0.0076 to -0.017, which is more than double the difference. That is, the effect of infection cases largely varies in the possible range of density and distance, and hence, their effects are significant. 
		\begin{figure}[H]
		\begin{center}
			\begin{tabular}{c}
				
				\begin{minipage}{0.5\hsize}
					\begin{center}
						\includegraphics[clip, width=80mm]{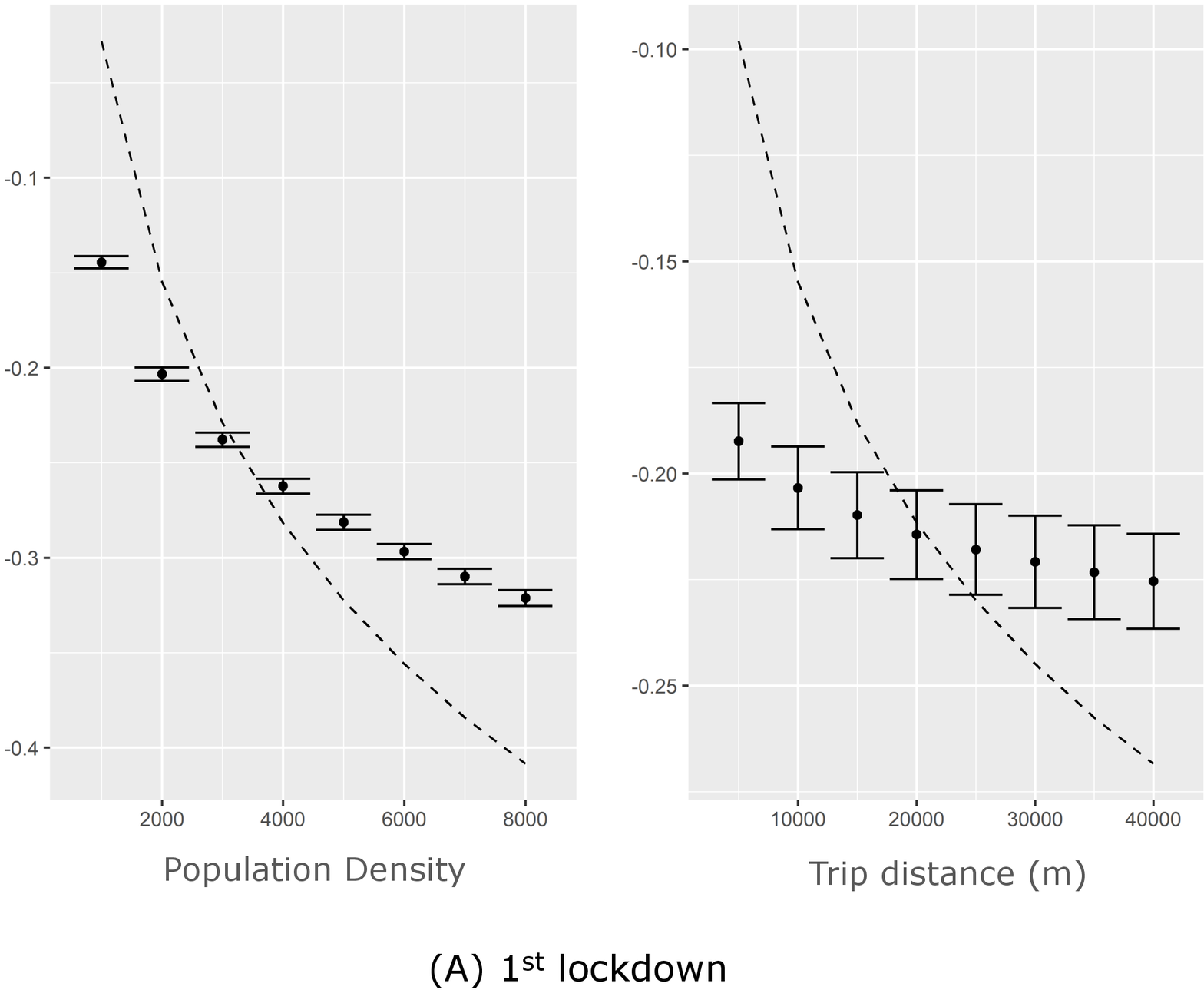}
					\end{center}
				\end{minipage}
				
				\begin{minipage}{0.5\hsize}
					\begin{center}
						\includegraphics[clip, width=80mm]{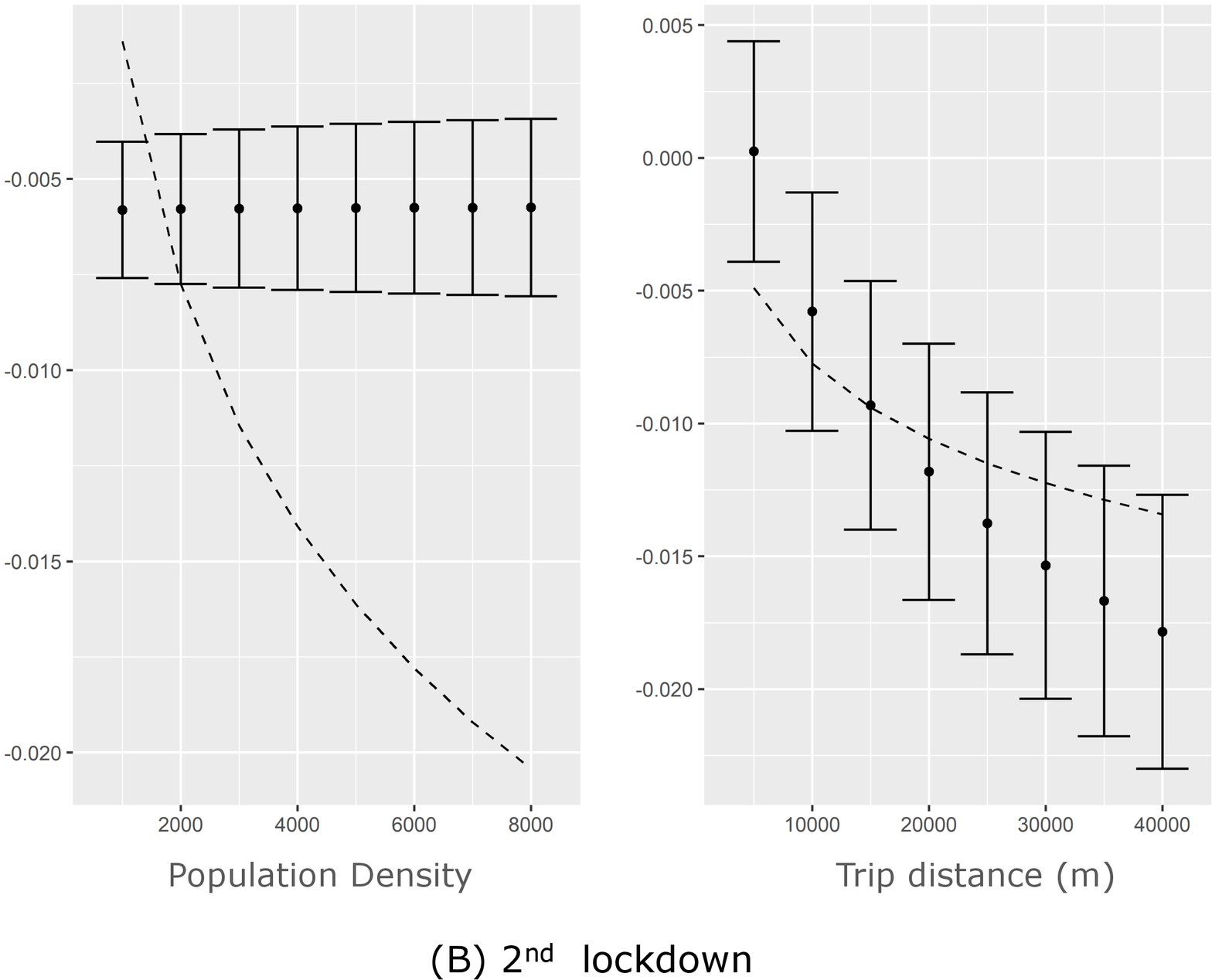}
					\end{center}
				\end{minipage}
				
			\end{tabular}
			\caption{{\bf Total marginal effects of the lockdowns: (A)First lockdown; (B) Second lockdown.} The total marginal effects of each lockdown are described by $ \beta_{0m}+\beta_{1m}\ln{Dist}+\beta_{2m}\ln{Dens}$, where $\beta_{0m}, \beta_{1m} $, and $ \beta_{2m} $ represent the coefficients of the single and cross terms of the mth lockdown in Model 2 of Table ~\ref{table1}. The dashed lines show the total effects of the IFC under the same conditions. Those with the first lockdown are multiplied by 20, but no scale adjustment is performed for the second lockdown.}
			\label{fig7}
		\end{center}
	\end{figure} 
	
	We also calculated the sensitivity of the total effects of the two lockdowns. Fig~\ref{fig7} shows that the total effect of the first lockdown also largely differs depending on the density and distance, as in IFC, although the lockdowns’ effects are slightly less sensitive than those of IFC. These results may be reasonable considering that, following ~\cite{watanabe2021japan}, lockdowns have two different effects: one is the information effect that lets people know the risks, and the other is the intervention effect to control people's mobility with legal or psychological coercion. 
	Here, if we assume that only the former effect depends on real risk factors such as population density or trip distance, these results imply the existence of the former. 
	The significance of the information effect was also shown by  ~\cite{watanabe2021japan}.
	Therefore, if governments are aware of the role of that effect in sending an effective message to people, they can use this political tool more effectively.
	
	However, the results for the second lockdown were weak and ambiguous, implying that it was less effective. Since Japan experienced at least two more lockdowns by July 31, 2021, after the data period, some additional examinations will be necessary to identify how that effect is still effective after the first lockdown. 
	
	\subsection{Dividing the data period}
	We divided the data into the early and late stages of the pandemic to investigate how people changed their attitude to the infection risks and the lockdowns during one year of COVID-19. The early stage is grouped from January 2020 to August 2020, while the latter is grouped from September 2020 to March 2021. 
	The estimation result from Eq~(\ref{eq:estimation}) is reported in Table~\ref{table2}. 
	\begin{table}[H]
			\centering
			\caption{
				{\bf Estimation results for earlier and later periods.}}
			\begin{tabular}{l c c c c}
				\hline
				& Model 1 & Model 2 & Model 3 & Model 4 \\
				\hline
				ln(IFC)            & $-0.0267^{***}$ & $-0.0124^{***}$ & $0.1219^{***}$  & $0.0489^{***}$  \\
				& $(0.0001)$      & $(0.0002)$      & $(0.0010)$      & $(0.0023)$      \\
				ln(LONG)           & $-0.0318^{***}$ & $-0.0141^{***}$ & $0.1041^{***}$  & $0.0253^{***}$  \\
				& $(0.0001)$      & $(0.0003)$      & $(0.0015)$      & $(0.0038)$      \\
				LD1                 & $-0.1620^{***}$ &                 & $0.5017^{***}$  &                 \\
				& $(0.0005)$      &                 & $(0.0059)$      &                 \\
				LD2                 &                 & $-0.0338^{***}$ &                 & $0.1983^{***}$  \\
				&                 & $(0.0003)$      &                 & $(0.0041)$      \\
				ln(IFC)$\times$ln(Dens)  &                 &                 & $-0.0136^{***}$ & $-0.0051^{***}$ \\
				&                 &                 & $(0.0001)$      & $(0.0002)$      \\
				ln(IFC)$\times$ln(Dist)  &                 &                 & $-0.0042^{***}$ & $-0.0026^{***}$ \\
				&                 &                 & $(0.0001)$      & $(0.0003)$      \\
				ln(LONG)$\times$ln(Dens) &                 &                 & $-0.0148^{***}$ & $-0.0052^{***}$ \\
				&                 &                 & $(0.0001)$      & $(0.0003)$      \\
				ln(LONG)$\times$ln(Dist) &                 &                 & $-0.0017^{***}$ & $-0.0002$       \\
				&                 &                 & $(0.0002)$      & $(0.0005)$      \\
				ln(Dens)$\times$LD1       &                 &                 & $-0.0640^{***}$ &                 \\
				&                 &                 & $(0.0004)$      &                 \\
				ln(Dist)$\times$LD1       &                 &                 & $-0.0186^{***}$ &                 \\
				&                 &                 & $(0.0008)$      &                 \\
				ln(Dens)$\times$LD2       &                 &                 &                 & $-0.0174^{***}$ \\
				&                 &                 &                 & $(0.0003)$      \\
				ln(Dist)$\times$LD2       &                 &                 &                 & $-0.0102^{***}$ \\
				&                 &                 &                 & $(0.0005)$      \\
				\hline 
				FE : OD             & YES           & YES           & YES           & YES           \\
				FE : date           & NO            & NO            & NO            & NO           \\
				\hline
				$R^2$               & $0.94$        & $0.96$        & $0.95$        & $0.96$        \\
				Adj. $R^2$          & $0.94$        & $0.96$        & $0.94$        & $0.96$        \\
				within.r.squared    & $0.18$        & $0.02$        & $0.25$        & $0.02$        \\
				Observations                & $5,218,383$     & $4,842,409$     & $5,218,383$     & $4,842,409$     \\
				\hline
			\end{tabular}
			\begin{flushleft}
				Notes: Figures in parentheses are cluster-robust standard errors by OD. $^{***}$, $^{**}$, and $^{*}$ denote statistical significance at the 0.1\% , 1\%, and 5\% level, respectively.
			\end{flushleft}
			\label{table2}
	\end{table}
	
	Although the sign of the effects is the same in both periods, the magnitudes of most coefficients are smaller in the later stage. This implies that people got used to COVID-19 and became less sensitive to short-term changes in infection cases and lockdowns.
	\footnote{In August and September, the second peak of infection finishes (see Fig~\ref{fig1}), and the Japanese government conducted ?go to travel campaign, which accelerated domestic travel to recover the economic situation from July 22, 2020.}
	
	\subsection{Mobility in weekend afternoon }
	Although this study has mainly focused on weekday mornings to investigate commuting trips, we also conducted the same analysis using the data from the weekend afternoon (14:00), which is considered to include trips for personal purposes such as leisure, shopping, and commuting. The results are reported in Table~\ref{table3} and are similar to weekday mornings; hence, most of our implications can be extended to various types of trips. However, this table may show several different tendencies for private trips. The positive cross term of IFC and Dens means that people do not refrain from trips to high-density places if the number of recent infection cases increases. This may be because some people change their travel plans in advance and quit distant trips based on the number of infected people. We need to implement further estimations to investigate these counterintuitive results. 
	
	\begin{table}[H]
			\centering
			\caption{
				{\bf Estimation results for the weekend afternoon.}}
			\begin{tabular}{l c c c c}
				\hline
				& Model 1 & Model 2 & Model 3 & Model 4 \\
				\hline
				ln(IFC)            & $-0.0106^{***}$ & $0.0741^{***}$  & $-0.0489^{***}$ & $0.0352^{***}$  \\
				& $(0.0001)$      & $(0.0017)$      & $(0.0005)$      & $(0.0017)$      \\
				ln(LONG)           & $0.0193^{***}$  & $0.0558^{***}$  & $-0.0327^{***}$ & $-0.0224^{***}$ \\
				& $(0.0002)$      & $(0.0031)$      & $(0.0004)$      & $(0.0030)$      \\
				LD1                 & $-0.2872^{***}$ & $1.8686^{***}$  &                 &                 \\
				& $(0.0015)$      & $(0.0161)$      &                 &                 \\
				LD2                 & $0.0027^{***}$  & $0.0339^{***}$  &                 &                 \\
				& $(0.0004)$      & $(0.0055)$      &                 &                 \\
				ln(IFC)$\times$ln(Dens)  &                 & $-0.0072^{***}$ &                 & $-0.0056^{***}$ \\
				&                 & $(0.0001)$      &                 & $(0.0001)$      \\
				ln(IFC)$\times$ln(Dist)  &                 & $-0.0031^{***}$ &                 & $-0.0031^{***}$ \\
				&                 & $(0.0002)$      &                 & $(0.0002)$      \\
				ln(LONG)$\times$ln(Dens) &                 & $0.0101^{***}$  &                 & $0.0092^{***}$  \\
				&                 & $(0.0002)$      &                 & $(0.0002)$      \\
				ln(LONG)$\times$ln(Dist) &                 & $-0.0122^{***}$ &                 & $-0.0071^{***}$ \\
				&                 & $(0.0004)$      &                 & $(0.0004)$      \\
				ln(Dens)$\times$LD1       &                 & $-0.1668^{***}$ &                 & $-0.1638^{***}$ \\
				&                 & $(0.0010)$      &                 & $(0.0010)$      \\
				ln(Dist)$\times$LD1       &                 & $-0.1007^{***}$ &                 & $-0.1033^{***}$ \\
				&                 & $(0.0019)$      &                 & $(0.0019)$      \\
				ln(Dens)$\times$LD2       &                 & $0.0052^{***}$  &                 & $-0.0010^{**}$  \\
				&                 & $(0.0004)$      &                 & $(0.0004)$      \\
				ln(Dist)$\times$LD2       &                 & $-0.0084^{***}$ &                 & $-0.0048^{***}$ \\
				&                 & $(0.0007)$      &                 & $(0.0007)$      \\
				\hline 
				FE : OD             & YES           & YES           & YES           & YES           \\
				FE : date           & NO            & NO            & YES           & YES           \\
				\hline
				$R^2$               & $0.92$        & $0.92$        & $0.92$        & $0.93$        \\
				Adj. $R^2$          & $0.91$        & $0.92$        & $0.92$        & $0.92$        \\
				within.r.squared    & $0.07$        & $0.12$        & $0.00$        & $0.06$        \\
				Sigma               & $0.28$        & $0.27$        & $0.27$        & $0.27$        \\
				Observations                & $6,206,073$     & $6,206,073$     & $6,206,073$     & $6,206,073$     \\
				\hline
			\end{tabular}
			\begin{flushleft}
				Notes: Figures in parentheses are cluster-robust standard errors by OD. $^{***}$, $^{**}$, and $^{*}$ denote statistical significance at the 0.1\% , 1\%, and 5\% level, respectively.
			\end{flushleft}
			\label{table3}
	\end{table}

	\section{Conclusion}
	We investigated how the effects of infection risks and voluntary lockdowns on mobility differ by distance and density using mobility data with detailed spatial characteristics, and the following three main results were obtained. The first is the mobility reduction caused by infection cases.
	People decrease their mobility after getting information about infection cases.
	The reduction of mobility increases with the population density of destination and the distance of the trip.
	This means people avoid the risk of infection considering the spatial characteristics of trips.
	Second is the information effect of the non-compulsory lockdowns.
	People avoided congested places and long-distance trips during lockdowns 
	as in the case of an increase in infection cases, which provides evidence that the information effect of the non-compulsory lockdowns alerts people to the infection risks.
	Third is the change in reaction of people during one year.
	Our results show that people react less sensitively to infection cases in the later periods (September 2020–March 2021) compared to the early periods (January 2020–July 2021), and the second lockdown was less effective than the first.

\end{document}